\documentstyle[11pt,aaspp4,psfig]{article}

\def\lesssim{\ifmmode {\,\mathbin{\lower 3pt\hbox   %< or of order
    {$\,\rlap{\raise 5pt\hbox{$\char'074$}}\mathchar"7218\,$}}}
    \else {${\mathbin{\lower 3pt\hbox
    {$\rlap{\raise 5pt\hbox{$\char'074$}}\mathchar"7218\,$}}}
    $}\fi}
\def\gtrsim{\ifmmode {\mathbin{\lower 3pt\hbox   %> or of order
    {$\,\rlap{\raise 5pt\hbox{$\char'076$}}\mathchar"7218\,$}}}
    \else {${\mathbin{\lower 3pt\hbox
    {$\rlap{\raise 5pt\hbox{$\char'076$}}\mathchar"7218\,$}}}
    $}\fi}

\lefthead{D.\,E. Holz}
\righthead{Lensing and high-$z$ supernova surveys}

\begin{document}

\title{Lensing and high-$z$ supernova surveys}

\author{Daniel E.\ Holz}
\affil{Enrico Fermi Institute and Department of Physics\\
       University of Chicago\\
       5640 South Ellis Avenue, Chicago, IL 60637-1433\\
       deholz@rainbow.uchicago.edu}
\authoremail{deholz@rainbow.uchicago.edu}

\begin{abstract}

Gravitational lensing causes the distribution of observed
brightnesses of standard candles at a given redshift to be
highly non-gaussian.
The distribution is strongly, and
asymmetrically, peaked at a value less than the expected value in a
homogeneous Robertson-Walker universe. Therefore, given any small
sample of observations in an inhomogeneous universe, the most likely
observed luminosity is at flux values less than the
Robertson-Walker value. This paper explores the impact of this
systematic error due to lensing upon surveys predicated on
measuring standard candle brightnesses.
We re-analyze recent results from the high-$z$
supernova team (\cite{Riess98}), both when most of the
matter in the universe is in the form of compact objects (represented
by the empty-beam expression, corresponding to the maximal case
of lensing), and when
the matter is continuously distributed in galaxies.
We find that the best-fit model remains
unchanged (at $\Omega_{m}=0$, $\Omega_\Lambda=0.45$), but the
confidence contours change size and shape, becoming larger (and thus
allowing a broader range of parameter space) and dropping towards
higher values of matter density, $\Omega_{m}$
(or correspondingly, lower values of the cosmological constant,
$\Omega_\Lambda$). These effects are slight when the matter is
continuously distributed. However, the effects become considerably
more important if most of the matter is in compact objects.
For example, neglecting lensing, the
$\Omega_{m}=0.5$, $\Omega_\Lambda=0.5$ model is more than $2\sigma$ away
from the best fit. In the empty-beam analysis, this cosmology
is at $1\sigma$.
\end{abstract}


\section{INTRODUCTION}

Recently there has been great activity in determining cosmological
parameters based on the observations of type Ia supernovae at high
redshifts (\cite{Perlmutter98}, 1997; \cite{Riess98}; \cite{Schmidt98}).
The peak flux
of these supernovae is thought to be known to within $0.15\ \mbox{mag}$
(\cite{Hamuy96}; \cite{RPK96}),
making them excellent standard candles with which to measure the
luminosity distance--redshift relation. As this
relation is dependent upon the cosmological parameters
($H_0$, $\Omega_{m}$, $\Omega_\Lambda$), it is possible to infer the
values of these parameters from supernova Ia observations.
Two independent groups (\cite{Perlmutter98}; \cite{Schmidt98})
have been pursuing
cosmology via supernova surveys, and
preliminary results argue for a low mass density (\cite{Perlmutter98};
\cite{Schmidt98}), and a nonzero cosmological constant (\cite{Riess98}).

The apparent brightness of a standard candle, at a given redshift, is
a function not only of the Robertson-Walker model describing the
universe, but also of the distribution of matter within the
universe. To date, the cosmology results in the literature derived
from observed supernova peak brightnesses are based upon the
assumption that the matter in the universe is homogeneously
distributed. An important question is to what extent the matter
inhomogeneities we see (in the form of galaxies, stars, MACHOs, etc.)
affect one's ability to draw conclusions from the data.

It is well known that weak gravitational lensing can impact the
degree to which high-redshift supernovae can be considered
standard candles, contributing
``errors'' on the order of $0.05\ \mbox{mag}$ by redshift of one for
various
cosmologies (\cite{KVB95}; \cite{Frieman97}; \cite{WCXO97};
\cite{HW98}; \cite{Kantowski98a}, 1998b).  These errors
can be beaten down through statistics, since, given a large enough
sample of supernovae at a given redshift, the average flux of the
supernovae should be representative of the true flux in a pure
Robertson-Walker universe.  Work has also
been done analyzing the high-magnification tail of the lensing
distribution, and the information this yields about the distribution
of mass inhomogeneities (\cite{SW87}).  In contrast,
the present paper examines the
low-magnification part of the lensing distribution, and explores the
systematic errors due to the non-gaussian peak. A
similar study has been undertaken by Kantowski (1998a, 1998b), examining
fits to analytic distance--redshift relations from Swiss-Cheese model
universes.
In this paper we utilize a newly-developed model to determine lensing
statistics (\cite{HW98}). With this model we are able to re-analyze current
results from supernova surveys, both in the case of the maximal (empty-beam)
lensing effects, and in the case where matter is distributed continuously
in galaxies.\footnote{The
results discussed in this paper
for the case of ``standard candles'' apply equally well to
``standard rulers''.
For example, the work of Guerra \& Daly (1998) utilizes
double radio sources to measure cosmological parameters, arriving at results
similar to those of the supernova groups. Lensing causes apparent
lengths to appear systematically shorter,
and thus engenders similar effects to those in the supernova case.}

\section{DISTANCE--REDSHIFT RELATIONS}

The luminosity of an image, as a
function of redshift, is related to the angular diameter distance to
the source generating the image.
Two common angular diameter distance--redshift relations are given by
the
filled and empty-beam expressions (\cite{DR72}, 1973; \cite{FFKT92}),
where in the filled-beam case
the line-of-sight to the source traverses mass of exactly
the Robertson-Walker density, while in the empty-beam case the beam
encounters no curvature (i.e., passes through vacuum, far from all
matter distributions).
Current analyses of high-$z$ supernova data use filled-beam expressions to
infer the physical distances of the sources from their observed
apparent brightnesses (\cite{Perlmutter98}; \cite{Riess98}).
When calculating how far a source is, based upon
the brightness of its image, it is therefore assumed that the photon
beams pass through exactly the Robertson-Walker mass density. If,
for example, the photon beams avoid most of the matter,
then a more accurate description would be the empty-beam
expressions. By
assuming a filled-beam expression in this case, one takes an
image dimmed because of the lack of matter in the
beam, and concludes from the
observed brightness of this image that the source is further away than
it really is.
With increasing redshift, the differences between the
filled and empty-beam brightnesses increase.  In this way, evidence
of an inhomogeneous universe might be mistaken for evidence of an
accelerating one.

The filled and empty-beam distance--redshift relations
reduce to the same form for low $z$, and
thus measurements of cosmological parameters
from supernovae with statistical weight at lower redshifts will
not be affected by lensing. As one moves to higher
redshifts, however, the differences between the two expressions
can be dramatic. For example, at $z=1$
a standard candle described by the filled-beam expression
in a smooth Robertson-Walker universe, with $\Omega_{m}=0.5$,
$\Omega_\Lambda=0.5$, will have the same apparent brightness as a
standard candle described by
the empty-beam expression in an inhomogeneous universe, with $\Omega_{m}=0.5$,
$\Omega_\Lambda=0$.
Therefore, based solely upon observations of standard candles
at both low redshifts and at
a single high redshift, it would be
impossible to conclude whether dimming
was due to lensing effects or a nonzero cosmological constant.
Knowledge of the distance--redshift curve
at a range of intermediate to high redshifts is thus crucial.

\section{MAGNIFICATION DISTRIBUTIONS}

We generate magnification distributions utilizing a recently-developed
method to determine lensing statistics in inhomogeneous universes
(\cite{HW98}).
In brief, the
method arrives at statistical lensing information by combining
aspects from ray-tracing and Swiss-Cheese model numerical approaches.
The universe is decomposed into
comoving spherical regions, with arbitrary mass inhomogeneities allowed within
each region. Statistics are developed by considering many
random rays in a Monte-Carlo fashion, and integrating the geodesic
deviation equation along each ray in turn. It should be emphasized that
this method calculates the lensing effects in full generality, treating
both weak and strong lensing effects automatically.

\begin{figure*}[t]
\hbox{\hskip 0.75truein
\psfig{file=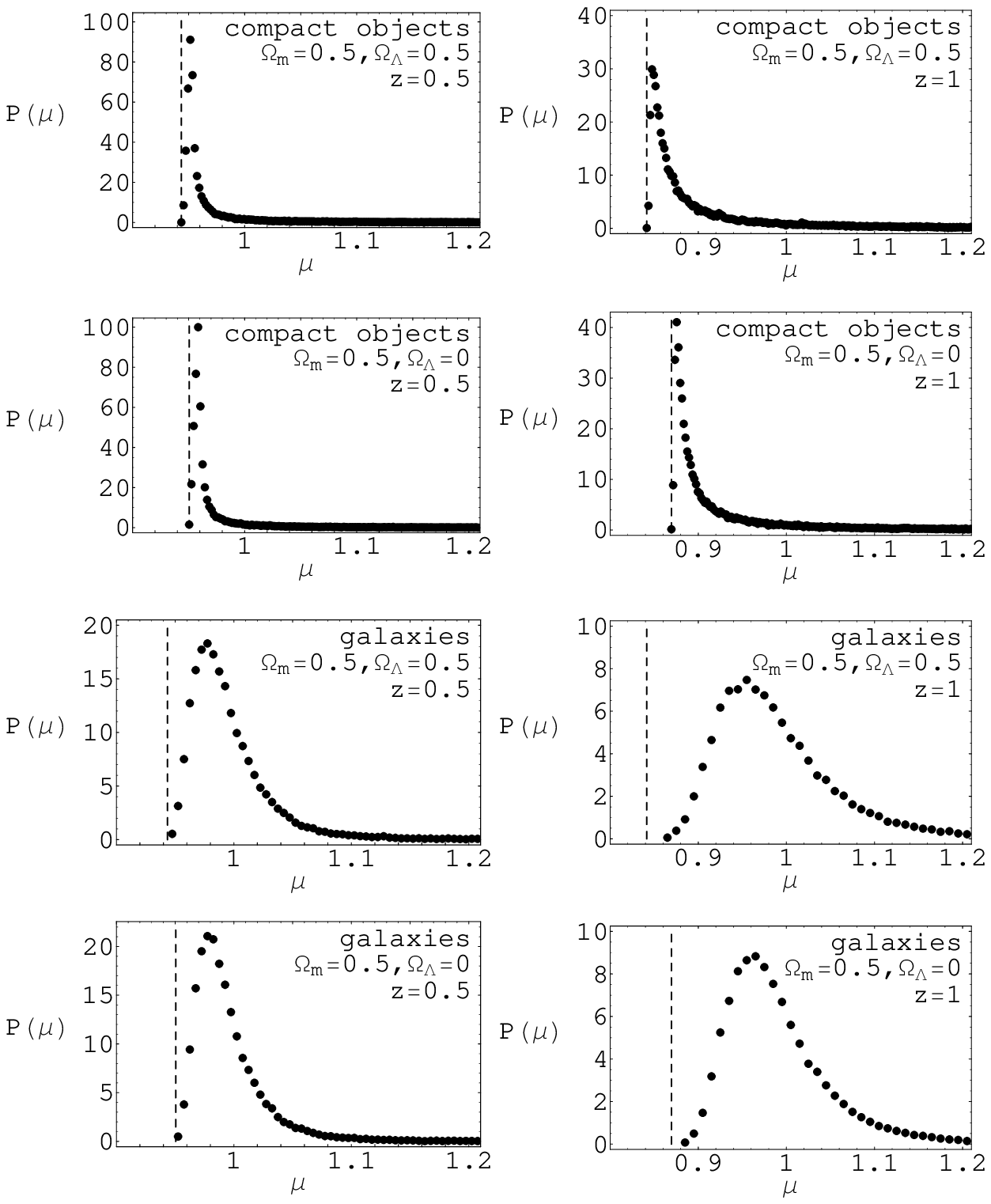,height=5.0truein,width=5.0truein}}
\caption[fig1]{
Probability distribution, $P(\mu)$, for supernova apparent
brightness, $\mu$, normalized so that $\mu=1$ corresponds to the
filled-beam, or homogeneous Robertson-Walker, value. The vertical
dashed lines are at the empty-beam value. The brightness
distributions are at $z=0.5$ and $z=1$, for the models as indicated.
The ``compact objects'' distribution is for the case where all of the matter
is in point masses. The ``galaxies'' distribution
is for matter in isothermal spheres with truncation radii $380\ \mbox{kpc}$.
}
\label{F:1}
\end{figure*}

Some representative magnification distributions are shown in
Figure~\ref{F:1}, for two different matter distributions.
The ``compact objects'' panels give the magnification distributions when most
of the matter in the universe is in highly condensed (point mass)
objects with masses
$\!\gtrsim0.01\ \mbox{M}_\odot$ (stars, MACHOs, etc.).
These results are
independent of the mass distribution and clustering of the point
masses (\cite{HW98}).
At the redshifts being considered, point mass lensing will in general
be dominated by a single lens encounter, and two images will be
generated.  The numerical method we utilize generates statistics for
uncorrelated photon beams, and thus does not identify the two images
associated with a given source (\cite{HW98}).  However, using analytic
expressions for the relative brightnesses of these images
(\cite{SEF92}), we are able to convert the magnification distribution
for the beams that have not passed through a caustic (which
correspond to the brighter image of a pair) into a magnification
distribution for the combined images.  The ``compact objects'' panels
of Figure~\ref{F:1} show this
combined magnification distribution.

For continuous matter distributions, as in
the case of isothermal galaxies, the incidence
of multiple imaging at the redshifts being considered is highly
improbable ($<\!\!0.2\%$ at $z=1$ (\cite{HMQ98})).
In this case strong lensing is unimportant, and weak lensing dominates
the results.
In the ``galaxies'' panels of Figure~\ref{F:1}, we have
taken all of the matter in the universe to be in galaxies, where a
galaxy is represented as a truncated isothermal
sphere of continuously distributed matter.
In this paper we take a galaxy number density of
$0.025\,h^3\ \mbox{Mpc}^{-3}$~(\cite{Geller97}), which,
for a given cosmology, fixes the mass of
the galaxies.
For example, taking $\Omega_{m}=0.5$ and $H_0=65\
\mbox{km}/\mbox{s}\mbox{\ Mpc}^{-1}$, the mass of the
galaxies is fixed at $8.5\times10^{12}\ \mbox{M}_\odot$.
If we then take the velocity dispersion of the galaxies to
be $220\ \mbox{km}/\mbox{s}$, the
physical truncation radius of the isothermal spheres is set at
$380\ \mbox{kpc}$.

A key feature of the magnification distributions plotted in
Figure~\ref{F:1} is that they are
non-gaussian.  The average of the distributions is given by the
Robertson-Walker (filled-beam) value ($\mu=1$).  In all cases,
however, the magnification distributions are strongly peaked at values
considerably less than this average. 
Another important characteristic of the distributions is that there is a lower
cutoff, given by the empty-beam value, to the possible observed
de-magnification. In general, the peak of the
magnification distribution
will lie somewhere in between the empty and filled-beam values.

With good statistics it may become possible to measure the
magnification distribution of the supernovae from observations, and
thus determine the matter distribution (\cite{Metcalf98};
\cite{WF98}).\footnote{
In this manner,
a cosmological MACHO experiment is possible.  For example, if most of
the matter is distributed in point masses, then the peak of the
magnification distribution will be very near the empty-beam value. In
this case, there will be little fluctuation to brightness values
dimmer than the mean, and considerably more fluctuation to brighter
values.  In addition, this asymmetry will grow with redshift in a well
defined manner. It is to be stressed that these MACHO detections are
possible even without strongly magnified supernovae or time-dependent
lensing effects.}
In the case of low statistics,
as is found in the high-$z$ supernova surveys, the likelihood of
evenly sampling the probability distribution is low.  One therefore
would expect these surveys to find a ``mean'' in rough agreement with
the (de-magnified) {\em peak} of the distribution, rather than the
average.
As the number of data points at a given redshift increases
($\gtrsim50$), the distribution of the average of the data points
approaches a gaussian distribution, centered about the average value
of the original distribution.  However, for the smaller high-$z$ data
samples currently available, the distributions of the averages retain
the highly non-gaussian form of the original (single data point)
distributions.\footnote{
For example, in the case of an $\Omega_m=0.5$, $\Omega_\Lambda=0$
model, with matter distributed in isothermal galaxies, we find
that the distribution of the average magnification of 10 supernovae
at redshift of $1/2$ is peaked
at a magnification halfway between the mode of the single
supernova distribution (shown in Fig.~\ref{F:1}) and the average of
the distribution ($\mu=1$). With matter in compact objects,
the peak of the 10-supernova distribution is
found to be $1/10$ of the way between the peak of the single data
point magnification distribution and the average.
}
Therefore, in what follows we take the magnification distributions to be
approximated by the peaks of the single data point distributions.
In the case of compact objects,
the distributions are sharply peaked
very near the empty-beam limits, and therefore the empty-beam
values are excellent approximations to the magnification distributions.
For continuous
matter distributions, such as extended galactic structures, the
peaks of the distributions are still considerably de-magnified from the
filled-beam value: The peak always falls in between the empty and
filled-beam values. Therefore, in
the case of extended matter distributions, doing an analysis with the
empty-beam expressions in addition to the filled-beam ones serves to
bracket the possible range of lensing effects.  In
the following section we analyze a sample high-$z$ supernova data,
fitting to both the empty-beam distance-redshift relation and
the peaks of the ``galaxies'' magnification distributions,
as well as the more traditional filled-beam distance-redshift relations.

\section{APPLICATION TO SUPERNOVA DATA}

We use a sample of supernovae as a test bed to determine
the qualitative effects of lensing on high-$z$ supernova
surveys.
To this
end, we take distance data for a total of 37 supernovae from the
high-$z$ supernova team: 27 supernovae at $0.01\lesssim z\lesssim0.1$
(\cite{Hamuy96}), and ten higher-redshift supernovae reported in Riess
et al.\ (1998a) (and analyzed using the MLCS method of Riess, Press,
\& Kirshner (1996)), including SN97ck at $z=0.97$
(\cite{Garnavich98}).\footnote{We have repeated the analysis
with the inclusion of the ``snapshot'' data
of Riess et al.\ (1998a) (which do not possess complete light curves),
and the results are very similar to those presented in this paper.}

For the low-redshift sample, we take the distance errors
listed in Table~10 of~Riess et al.\ (1998a),
and the dispersion in
host galaxy redshifts (due to peculiar velocities and
other uncertainties) to be $200\ \mbox{km}/\mbox{s}$.
For the high-$z$ sample, we take the
errors listed in Riess et al.\ (1998a).
We find that the fits are not sensitive to the
particular error values, in agreement with Riess
et al.\ (1998a).

We fix the Hubble constant to be $65\
\mbox{km}/\mbox{s}\mbox{\ Mpc}^{-1}$, in accord with the determination
of Riess et al.\ (1998a). As $H_0$ is determined from supernovae (or other
methods) at low redshifts, gravitational lensing is not expected
to affect this result. We stress that all of
the results discussed in this paper are independent of the
value of $H_0$.

\begin{figure*}[t]
\hbox{\hskip .5truein
\psfig{file=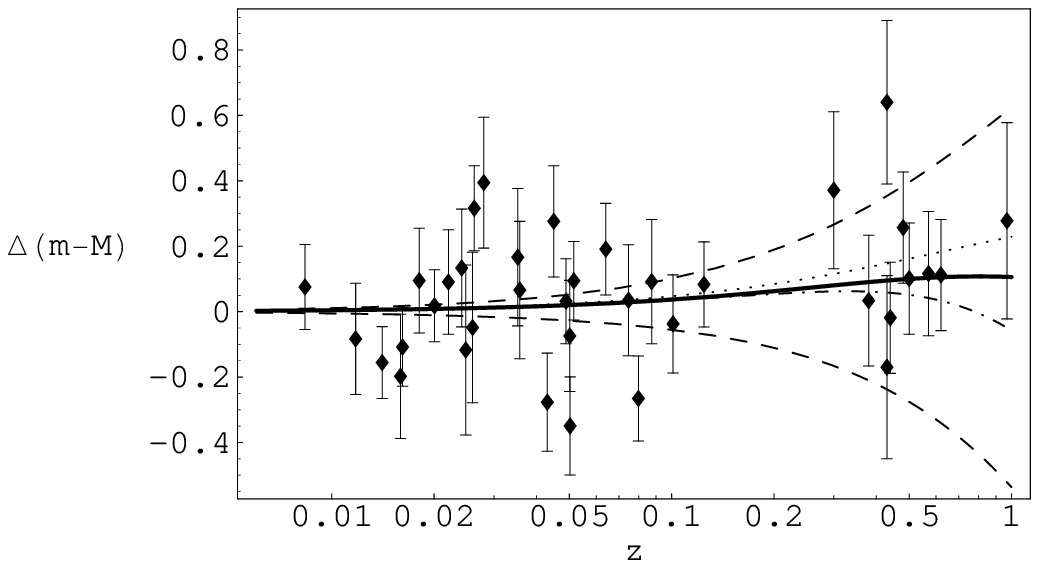}}
\caption[fig2]{
Residual Hubble diagram (the values of
the $\Omega_{m}=\Omega_\Lambda=0$ model have been subtracted out), with
differential distance modulus plotted against redshift, for the supernova
data sample and a number of theoretical distance-redshift relations. 
The top (dashed) curve is for an $\Omega_{m}=0$, $\Omega_\Lambda=1$ model.
The bottom (dashed) curve is for an $\Omega_{m}=1$, $\Omega_\Lambda=0$
model. 
The dotted curve is the best fit to the data, corresponding to
$\Omega_{m}=0$, $\Omega_\Lambda=0.45$.
The solid curve is the empty-beam distance--redshift
relation for $\Omega_{m}=0.4$, $\Omega_\Lambda=0.6$. The dot-dashed
curve is the filled-beam relation for
the same parameters. Note that the empty-beam $\Omega_{m}=0.4$,
$\Omega_\Lambda=0.6$ case is essentially as good as the
best fit ($\chi_\nu^2=1.17$),
while the
corresponding filled-beam case is well over $1\sigma$
away ($\chi_\nu^2=1.24$).
}
\label{F:2}
\end{figure*}

In parallel with \S\,4.1 of Riess et al. (1998a), we do a two
parameter ($\Omega_{m}$, $\Omega_\Lambda$) minimum $\chi^2$ fit,
neglecting regions with $\Omega_{m}<0$, and other unphysical regions (``no
big bang'' regions (\cite{CPT92})).
The residual Hubble-diagram (where the
$\Omega_{m}=\Omega_\Lambda=0$ magnitudes have been subtracted out)
for the data set is shown in Figure~\ref{F:2}, plotted against the best-fit
(minimum $\chi^2$) curve, as well as some fiducial curves for reference.
The
best fit curve is a model with $\Omega_{m}=0$, $\Omega_\Lambda=0.45$
($\chi_\nu^2=1.15$, for 35 degrees of freedom).
This is in good agreement with the result
of Riess et al.\ (1998a) ($\Omega_{m}=0$, $\Omega_\Lambda=0.48$
($\chi_\nu^2=1.17$)),
which comes from an identical set of supernovae, but with updated MLCS
values (\cite{Riess98b}). Both empty and filled-beam
models find this value (which is identical in each case) as their best
fit. Note that the data
points
tend toward values above the axis, indicating a nonzero
cosmological constant, regardless of lensing. Also note
that the models become most clearly separated at high-$z$, and therefore
the discretionary power lies in the few highest-redshift
supernovae. This can readily be seen by the striking contrast between contours
from Perlmutter et al.\ (1997) and those from Perlmutter et al.\ (1998),
where the latter paper includes the addition of a single supernova
at $z=0.83$.
The more sensitive the fits are to the highest-$z$ supernovae,
the more lensing effects can come into play.

\begin{figure*}[t]
\hbox{\hskip -0.5truein
\psfig{file=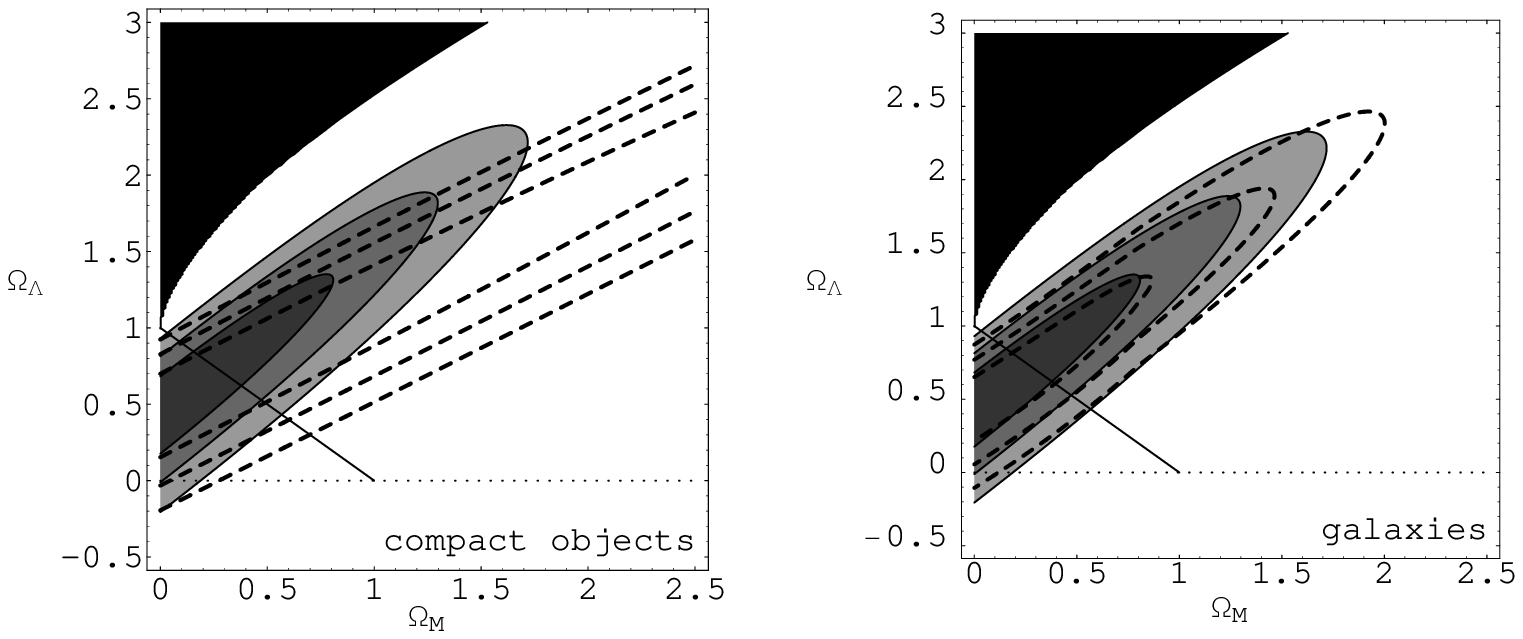}}
\caption[fig3]{
Contours of constant $\chi^2$ in
($\Omega_{m}$, $\Omega_\Lambda$) parameter space, for the SNe data set
of Fig.~\ref{F:2}. The background shaded contours correspond to the
68\%, 95\%, and 99.7\% joint confidence intervals in the filled-beam
case. The dotted line represents $\Lambda=0$ cosmologies,
while the diagonal line represents
$\Omega_{m}+\Omega_\Lambda=1$ cosmologies.
The blackened region at the upper left is excluded (no big
bang).
The ``compact objects'' panel includes the confidence intervals
(given by the foreground dashed lines) for the empty-beam case.
The ``galaxies'' panel includes the confidence intervals (given by the
foreground dashed lines) for the
case where matter is assumed to be continuously distributed in
isothermal galaxies, and the data
is fit to the peaks of the magnification distributions.
}
\label{F:3}
\end{figure*}

Although the empty and filled-beam cases agree on a best-fit model, in
neither case is the fit particularly tight. Therefore it is particularly
informative to consider likelihood contours, as discussed in Riess
et al.\ (1998a).
By integrating over successive
regions of ($\Omega_{m}$, $\Omega_\Lambda$) phase space, we are able to
determine the values of $\chi^2$ corresponding to the 68\%, 95\%, and
99.7\% confidence boundaries (representing $1\sigma$, $2\sigma$, and
$3\sigma$ regions of fit, respectively).
Figure~\ref{F:3} shows contours of
constant $\chi^2$ representative of these $1\sigma$, $2\sigma$, and
$3\sigma$ confidence intervals. In both panels the background shaded contours
give the standard filled-beam results.
The ``compact objects'' panel includes the respective contours
when the data is fit to empty-beam
expressions, corresponding to the maximal lensing case.
The ``galaxies'' panel fits the data to the magnification distributions
for isothermal galaxies, of the form shown in Figure~\ref{F:1}.
It is possible to estimate the value and width of the
peaks of these magnification distributions.
However, computing a magnification distribution for every point
in parameter space, and at each redshift for which there exists data,
is numerically prohibitively expensive. For our purposes we have
computed magnification distributions for 15 different models, at 4
different redshifts, and interpolated to arrive at peak and width
values for the magnification distributions in general. The confidence
contours of the ``galaxies'' panel are for the
case where these interpolated peak and width values have been utilized
to fit to the data.

All of the contours of Figure~\ref{F:3}
are similar near the $\Omega_{m}=0$ axis,
where there is little matter to cause lensing.
As one progresses to models with more significant matter content,
with the matter primarily in the form
of compact objects (the empty-beam case), the
contours remain wider than their filled-beam counterparts,
closing off at much larger
$\Omega_{m}$ and $\Omega_\Lambda$ values.
In this case the inclusion of lensing broadens the class of consistent
cosmological models.
Furthermore, the empty-beam contours drift downwards: the empty-beam fits
prefer higher values of $\Omega_{m}$, and lower values of $\Omega_\Lambda$.
These results appear to agree with preliminary results from the Supernova
Cosmology Project (\cite{Aldering98}).
Although these statements remain true when matter is continuously
distributed in galaxies, as can be seen from Figure~\ref{F:3}
the effects in this latter case are greatly reduced.

\section{CONCLUSIONS}

We have argued that lensing will systematically
skew the peak of
the apparent brightness distribution of supernovae away from the
filled-beam value and towards the
empty-beam value.
Based upon our results from a limited number of supernovae data points,
we can make
some qualitative statements regarding the impact of this systematic
effect on the determination of cosmological parameters from high-$z$ supernova
surveys.
For universes with little or no
$\Omega_{m}$, the effects of lensing are slight.
As the current data samples seem
to favor vacuum models, lensing will not generally affect their
best fits. However, the error ellipses undergo significant changes due
to the inclusion of lensing, favoring models with lower values of
$\Omega_\Lambda$ and higher values of $\Omega_{m}$.
If most of the matter in the universe is in the form of compact
objects, the effects of lensing can be dramatic.
For example, the empty-beam best fit
flat model ($\Omega_{m}+\Omega_\Lambda=1$) has $\Omega_m=0.32$
($\chi^2=1.155$). This fit is essentially as good as the overall
best fit ($\Omega_{m}=0$, $\Omega_\Lambda=0.45$ ($\chi^2=1.152$)).
The filled-beam best fit flat model
has $\sim\!25\%$ less matter ($\Omega_{m}=0.26$),
and is a slightly worse fit ($\chi^2=1.161$).
If most of the matter is continuously distributed, however, the effects of
lensing are greatly reduced.

Currently the true best fit to the data finds negative values for
$\Omega_{m}$.  As we neglect $\Omega_{m}<0$ on physical grounds, the
likelihood contours are squashed up against the $\Omega_{m}=0$ axis,
minimizing the effects of lensing.
Should future data favor a positive value of
$\Omega_{m}$, lensing can be expected to have a greater impact
on the analysis.

\acknowledgements

The author wishes to thank Don Lamb, Adam Riess, Mike Turner,
and especially Bob
Wald for valuable discussions in the course of this work. The author
also wishes to acknowledge Warner Miller for the initial stimulus to
undertake this project.  This work was supported by NSF grant PHY
95-14726.

\end{document}